\documentclass{article}
\usepackage{frascatiphys}
\usepackage{graphicx}
\def\kevc1{\ifmmode\mathrm{\ keV/{\mit c}}
          \else$\mathrm{\ keV/{\mit c}}$\fi}
\def\Mevc1{\ifmmode\mathrm{\ MeV/{\mit c}}
          \else$\mathrm{\ MeV/{\mit c}}$\fi}
\def\gevc1{\ifmmode\mathrm{\ GeV/{\mit c}}
          \else$\mathrm{\ GeV/{\mit c}}$\fi}
\def\kevc2{\ifmmode\mathrm{\ keV/{\mit c}^2}
          \else$\mathrm{\ keV/{\mit c}^2}$\fi}
\def\Mevc2{\ifmmode\mathrm{\ MeV/{\mit c}^2}
          \else$\mathrm{\ MeV/{\mit c}^2}$\fi}
\def\Gevc2{\ifmmode\mathrm{\ GeV/{\mit c}^2}
          \else$\mathrm{\ GeV/{\mit c}^2}$\fi}
\def\Gev2c2{\ifmmode\mathrm{\ GeV^2/{\mit c}^2}
          \else$\mathrm{\ GeV^2/{\mit c}^2}$\fi}

\def\ubar{\ifmmode\mathrm{\overline {u}}
          \else$\mathrm{\overline{u}}$\fi}
\def\dbar{\ifmmode\mathrm{\overline {d}}
          \else$\mathrm{\overline{d}}$\fi}
\def\sbar{\ifmmode\mathrm{\overline {s}}
          \else$\mathrm{\overline{s}}$\fi}
\def\cbar{\ifmmode\mathrm{\overline {c}}
          \else$\mathrm{\overline{c}}$\fi}
\def\bbar{\ifmmode\mathrm{\overline {b}}
          \else$\mathrm{\overline{b}}$\fi}
\def\tbar{\ifmmode\mathrm{\overline {t}}
          \else$\mathrm{\overline{t}}$\fi}
\def\qbar{\ifmmode\mathrm{\overline {q}}
          \else$\mathrm{\overline{q}}$\fi}

\def\uq{\ifmmode\mathrm{u}
          \else$\mathrm{u}$\fi}
\def\dq{\ifmmode\mathrm{d}
          \else$\mathrm{d}$\fi}
\def\sq{\ifmmode\mathrm{s}
          \else$\mathrm{s}$\fi}
\def\cq{\ifmmode\mathrm{c}
          \else$\mathrm{c}$\fi}
\def\bq{\ifmmode\mathrm{b}
          \else$\mathrm{b}$\fi}
\def\tq{\ifmmode\mathrm{t}
          \else$\mathrm{t}$\fi}
\def\qq{\ifmmode\mathrm{q}
          \else$\mathrm{q}$\fi}

 \def\Pgg{\ifmmode\mathrm{\gamma}
          \else$\mathrm{\gamma}$\fi}
 \def\PW{\ifmmode\mathrm{W}
         \else$\mathrm{W }$\fi}
 \def\PWp{\ifmmode\mathrm{W^+}
          \else$\mathrm{W^+}$\fi}
 \def\PWpm{\ifmmode\mathrm{W^{\pm}}
          \else$\mathrm{W^{\pm}}$\fi}
 \def\PWm{\ifmmode\mathrm{W^-}
          \else$\mathrm{W^-}$\fi}
 \def\PZz{\ifmmode\mathrm{Z^0}
          \else$\mathrm{Z^0}$\fi}
 \def\PHz{\ifmmode\mathrm{H^0}
          \else$\mathrm{H^0}$\fi}
 \def\PHpm{\ifmmode\mathrm{H^{\pm}}
           \else$\mathrm{H^{\pm}}$\fi}
 \def\PWR{\ifmmode\mathrm{W_R}
          \else$\mathrm{W_R}$\fi}
 \def\PWpr{\ifmmode\mathrm{W^{\prime}}
           \else$\mathrm{W^{\prime}}$\fi}
 \def\PZLR{\ifmmode\mathrm{Z_{LR}}
           \else$\mathrm{Z_{LR}}$\fi}
 \def\PZgc{\ifmmode\mathrm{Z_{\chi}}
           \else$\mathrm{Z_{\chi}}$\fi}
 \def\PZgy{\ifmmode\mathrm{Z_{\psi}}
           \else$\mathrm{Z_{\psi}}$\fi}
 \def\PZge{\ifmmode\mathrm{Z_{\eta}}
           \else$\mathrm{Z_{\eta}}$\fi}
 \def\PZi{\ifmmode\mathrm{Z_1}
          \else$\mathrm{Z_1}$\fi}
 \def\PAz{\ifmmode\mathrm{A^0}
          \else$\mathrm{A^0}$\fi}
 \def\Pgne{\ifmmode\mathrm{\nu_{e}}
           \else$\mathrm{\nu_{e}}$\fi}
 \def\Pagne{\ifmmode\mathrm{\overline{\nu_{e}}}
            \else$\mathrm{\overline{\nu_{e}}}$\fi}
 \def\Pgngm{\ifmmode\mathrm{\nu_{\mu}}
            \else$\mathrm{\nu_{\mu}}$\fi}
 \def\Pagngm{\ifmmode\mathrm{\overline{\nu}_{\mu}}
             \else$\mathrm{\overline{\nu}_{\mu}}$\fi}
 \def\Pgngt{\ifmmode\mathrm{\nu_{\tau}}
            \else$\mathrm{\nu_{\tau}}$\fi}
 \def\Pagngt{\ifmmode\mathrm{\overline{\nu}_{\tau}}
             \else$\mathrm{\overline{\nu}_{\tau}}$\fi}
 \def\Pe{\ifmmode\mathrm{e}
         \else$\mathrm{e}$\fi}
 \def\Pep{\ifmmode\mathrm{e^+}
          \else$\mathrm{e^+}$\fi}
 \def\Pem{\ifmmode\mathrm{e^-}
          \else$\mathrm{e^-}$\fi}
 \def\Pgm{\ifmmode\mathrm{\mu}
          \else$\mathrm{\mu}$\fi}
 \def\Pgmm{\ifmmode\mathrm{\mu^-}
           \else$\mathrm{\mu^-}$\fi}
 \def\Pgmp{\ifmmode\mathrm{\mu^+}
           \else$\mathrm{\mu^+}$\fi}
 \def\Pgt{\ifmmode\mathrm{\tau}
          \else$\mathrm{\tau}$\fi}
 \def\PLpm{\ifmmode\mathrm{L^{\pm}}
           \else$\mathrm{L^{\pm}}$\fi}
 \def\PLz{\ifmmode\mathrm{L^0}
          \else$\mathrm{L^0}$\fi}
 \def\PEz{\ifmmode\mathrm{E^0}
          \else$\mathrm{E^0}$\fi}
 \def\Pgp{\ifmmode\mathrm{\pi}
          \else$\mathrm{\pi }$\fi}
 \def\Pgpm{\ifmmode\mathrm{\pi^-}
           \else$\mathrm{\pi^-}$\fi}
 \def\Pgpp{\ifmmode\mathrm{\pi^+}
           \else$\mathrm{\pi^+}$\fi}
 \def\Pgppm{\ifmmode\mathrm{\pi^{\pm }}
            \else$\mathrm{\pi^{\pm }}$\fi}
 \def\Pgpz{\ifmmode\mathrm{\pi^0}
           \else$\mathrm{\pi^0 }$\fi}
 \def\Pgh{\ifmmode\mathrm{\eta}
          \else$\mathrm{\eta }$\fi}
 \def\Pgr{\ifmmode\mathrm{\rho(770)}
          \else$\mathrm{\rho(770)}$\fi}
 \def\Pgo{\ifmmode\mathrm{\omega(783)}
          \else$\mathrm{\omega(783)}$\fi}
 \def\Pghpr{\ifmmode\mathrm{\eta^{\prime}(958)}
            \else$\mathrm{\eta^{\prime}(958)}$\fi}
 \def\Pfz{\ifmmode\mathrm{f_0(980)}
          \else$\mathrm{f_0(980)}$\fi}
 \def\Paz{\ifmmode\mathrm{a_0(980)}
          \else$\mathrm{a_0(980)}$\fi}
 \def\Pgf{\ifmmode\mathrm{\phi(1020)}
          \else$\mathrm{\phi(1020)}$\fi}
 \def\Phia{\ifmmode\mathrm{h_1(1170)}
           \else$\mathrm{h_1(1170)}$\fi}
 \def\Pbi{\ifmmode\mathrm{b_1(1235)}
          \else$\mathrm{b_1(1235)}$\fi}
 \def\Pai{\ifmmode\mathrm{a_1(1260)}
          \else$\mathrm{a_1(1260)}$\fi}
 \def\Pfii{\ifmmode\mathrm{f_2(1270)}
           \else$\mathrm{f_2(1270)}$\fi}
 \def\Pfi{\ifmmode\mathrm{f_1(1285)}
          \else$\mathrm{f_1(1285)}$\fi}
 \def\Pgha{\ifmmode\mathrm{\eta(1295)}
           \else$\mathrm{\eta(1295)}$\fi}
 \def\Pgpa{\ifmmode\mathrm{\pi(1300)}
           \else$\mathrm{\pi(1300)}$\fi}
 \def\Paii{\ifmmode\mathrm{a_2(1320)}
           \else$\mathrm{a_2(1320)}$\fi}
 \def\Pgoa{\ifmmode\mathrm{\omega(1390)}
           \else$\mathrm{\omega(1390)}$\fi}
 \def\Pfza{\ifmmode\mathrm{f_0(1400)}
           \else$\mathrm{f_0(1400)}$\fi}
 \def\Pfia{\ifmmode\mathrm{f_1 (1390)}
           \else$\mathrm{f_1 (1390)}$\fi}
 \def\Pghb{\ifmmode\mathrm{\eta(1440)}
           \else$\mathrm{\eta(1440)}$\fi}
 \def\Pgra{\ifmmode\mathrm{\rho(1450)}
           \else$\mathrm{\rho(1450)}$\fi}
 \def\Pfib{\ifmmode\mathrm{f_1(1510)}
           \else$\mathrm{f_1(1510)}$\fi}
 \def\Pfiipr{\ifmmode\mathrm{f^{\prime}_2(1525)}
             \else$\mathrm{f^{\prime}_2(1525)}$\fi}
 \def\Pfzb{\ifmmode\mathrm{f_0(1590)}
           \else$\mathrm{f_0(1590)}$\fi}
 \def\Pgob{\ifmmode\mathrm{\omega(1600)}
           \else$\mathrm{\omega(1600)}$\fi}
 \def\Pgoiii{\ifmmode\mathrm{\omega_3(1670)}
             \else$\mathrm{\omega_3(1670)}$\fi}
 \def\Pgpii{\ifmmode\mathrm{\pi_2(1670)}
            \else$\mathrm{\pi_2(1670)}$\fi}
 \def\Pgfa{\ifmmode\mathrm{\phi(1680)}
           \else$\mathrm{\phi(1680)}$\fi}
 \def\Pgriii{\ifmmode\mathrm{\rho_3(1690)}
             \else$\mathrm{\rho_3(1690)}$\fi}
 \def\Pgrb{\ifmmode\mathrm{\rho(1700)}
           \else$\mathrm{\rho(1700)}$\fi}
 \def\Pfiia{\ifmmode\mathrm{f_2(1720)}
            \else$\mathrm{f_2(1720)}$\fi}
 \def\Pgfiii{\ifmmode\mathrm{\phi_3(1850)}
             \else$\mathrm{\phi_3(1850)}$\fi}
 \def\Pfiib{\ifmmode\mathrm{f_2(2010)}
            \else$\mathrm{f_2(2010)}$\fi}
 \def\Pfiv{\ifmmode\mathrm{f_4(2050)}
           \else$\mathrm{f_4(2050)}$\fi}
 \def\Pfiic{\ifmmode\mathrm{f_2(2300)}
            \else$\mathrm{f_2(2300)}$\fi}
 \def\Pfiid{\ifmmode\mathrm{f_2(2340)}
            \else$\mathrm{f_2(2340)}$\fi}
 \def\PK{\ifmmode\mathrm{K}
         \else$\mathrm{K}$\fi}
 \def\PKpm{\ifmmode\mathrm{K^{\pm}}
           \else$\mathrm{K^{\pm}}$\fi}
 \def\PKp{\ifmmode\mathrm{K^+}
          \else$\mathrm{K^+}$\fi}
 \def\PKm{\ifmmode\mathrm{K^-}
          \else$\mathrm{K^-}$\fi}
 \def\PKz{\ifmmode\mathrm{K^0}
          \else$\mathrm{K^0}$\fi}
 \def\PaKz{\ifmmode\mathrm{\overline{K^0}}
           \else$\mathrm{\overline{K^0}}$\fi}
 \def\PKgmiii{\ifmmode\mathrm{K_{\mu 3}}
              \else$\mathrm{K_{\mu 3}}$\fi}
 \def\PKeiii{\ifmmode\mathrm{K_{\rm e3}}
             \else$\mathrm{K_{\rm e3}}$\fi}
 \def\PKzS{\ifmmode\mathrm{K^0_{\rm S}}
           \else$\mathrm{K^0_{\rm S}}$\fi}
 \def\PKzL{\ifmmode\mathrm{K^0_{\rm L}}
           \else$\mathrm{K^0_{\rm L}}$\fi}
 \def\PKzgmiii{\ifmmode\mathrm{K^0_{\mu 3}}
               \else$\mathrm{K^0_{\mu 3}}$\fi}
 \def\PKzeiii{\ifmmode\mathrm{K^0_{{\rm e}3}}
              \else$\mathrm{K^0_{{\rm e}3}}$\fi}
 \def\PKst{\ifmmode\mathrm{K^{\ast}(892)}
           \else$\mathrm{K^{\ast}(892)}$\fi}
 \def\PKi{\ifmmode\mathrm{K_1(1270)}
          \else$\mathrm{K_1(1270)}$\fi}
 \def\PKsta{\ifmmode\mathrm{K^{\ast}(1370)}
            \else$\mathrm{K^{\ast}(1370)}$\fi}
 \def\PKia{\ifmmode\mathrm{K_1(1400)}
           \else$\mathrm{K_1(1400)}$\fi}
 \def\PKstz{\ifmmode\mathrm{K^{\ast}_0(1430)}
            \else$\mathrm{K^{\ast}_0(1430)}$\fi}
 \def\PKstii{\ifmmode\mathrm{K^{\ast}_2(1430)}
             \else$\mathrm{K^{\ast}_2(1430)}$\fi}
 \def\PKstb{\ifmmode\mathrm{K^{\ast}(1680)}
            \else$\mathrm{K^{\ast}(1680)}$\fi}
 \def\PKii{\ifmmode\mathrm{K_2(1770)}
           \else$\mathrm{K_2(1770)}$\fi}
 \def\PKstiii{\ifmmode\mathrm{K^{\ast}_3(1780)}
              \else$\mathrm{K^{\ast}_3(1780)}$\fi}
 \def\PKstiv{\ifmmode\mathrm{K^{\ast}_4(2045)}
             \else$\mathrm{K^{\ast}_4(2045)}$\fi}
 \def\PD{\ifmmode\mathrm{D}
           \else$\mathrm{D}$\fi}
 \def\PaD{\ifmmode\mathrm{\overline{ D}}
          \else${\mathrm{\overline D}}$\fi}
 \def\PDpm{\ifmmode\mathrm{D^{\pm}}
           \else$\mathrm{D^{\pm}}$\fi}
 \def\PDm{\ifmmode\mathrm{D^-}
          \else$\mathrm{D^-}$\fi}
 \def\PDp{\ifmmode\mathrm{D^+}
          \else$\mathrm{D^+}$\fi}
 \def\PDz{\ifmmode\mathrm{D^0}
          \else$\mathrm{D^0}$\fi}
 \def\PaDz{\ifmmode\mathrm{\overline{D^0}}
           \else$\mathrm{\overline{D^0}}$\fi}
 \def\PDstpm{\ifmmode{\mathrm{D}^{\ast}(2010)^{\pm}}
             \else$\mathrm{D}^{\ast}(2010)^{\pm}$\fi}
 \def\PDstp{\ifmmode{\mathrm{D}^{\ast+}}
             \else$\mathrm{D}^{\ast+}$\fi}
 \def\PDst{\ifmmode{\mathrm{D}^{\ast}}
             \else$\mathrm{D}^{\ast}$\fi}
 \def\PDstz{\ifmmode{\mathrm{D}^{\ast}(2010)^0}
            \else$\mathrm{D}^{\ast}(2010)^0$\fi}
 \def\PDiz{\ifmmode{\mathrm{D}_{1}(2420)^0}
           \else$\mathrm{D}_{1}(2420)^0$\fi}
 \def\PDstiiz{\ifmmode{\mathrm{D}^{\ast}_{2}(2460)^0}
              \else$\mathrm{D}^{\ast}_{2}(2460)^0$\fi}
 \def\PsDp{\ifmmode\mathrm{D_{s}^+}
           \else$\mathrm{D_{s}^+}$\fi}
 \def\PsDm{\ifmmode\mathrm{D_{s}^-}
           \else$\mathrm{D_{s}^-}$\fi}
 \def\PsDpm{\ifmmode\mathrm{D_{s}^{\pm}}
           \else$\mathrm{D_{s}^{\pm}}$\fi}
 \def\PsDst{\ifmmode\mathrm{D_{s}^{\ast}}
            \else$\mathrm{D_{s}^{\ast}}$\fi}
 \def\PsDipm{\ifmmode\mathrm{D_{s1}(2536)^{\pm}}
           \else$\mathrm{D_{s1}(2536)^{\pm}}$\fi}
 \def\PB{\ifmmode{\mathrm{B}}
          \else$\mathrm{B}$\fi}
 \def\PBp{\ifmmode{\mathrm{B}^{+}}
           \else$\mathrm{B}^{+}$\fi}
 \def\PBm{\ifmmode{\mathrm{B}^{-}}
           \else$\mathrm{B}^{-}$\fi}
 \def\PBpm{\ifmmode{\mathrm{B}^{\pm}}
            \else$\mathrm{B}^{\pm}$\fi}
 \def\PBz{\ifmmode{\mathrm{B}^0}
           \else$\mathrm{B}^0$\fi}
 \def\PbgL{\ifmmode{\mathrm{\Lambda}_b}
           \else$\mathrm{\Lambda}_b$\fi}
 \def\Pcgh{\ifmmode\mathrm{{\eta}_{c}(1S)}
           \else$\mathrm{{\eta}_{c}(1S)}$\fi}
 \def\PJgyy{\ifmmode\mathrm{J /\psi}
           \else$\mathrm{J /\psi}$\fi}
 \def\PJgy{\ifmmode\mathrm{J /\psi(1S)}
           \else$\mathrm{J /\psi(1S)}$\fi}
 \def\Pcgcz{\ifmmode\mathrm{{\chi}_{c0}(1P)}
            \else$\mathrm{{\chi}_{c0}(1P)}$\fi}
 \def\Pcgci{\ifmmode\mathrm{{\chi}_{c1}(1P)}
            \else$\mathrm{{\chi}_{c1}(1P)}$\fi}
 \def\Pcgcii{\ifmmode\mathrm{{\chi}_{c2}(1P)}
             \else$\mathrm{{\chi}_{c2}(1P)}$\fi}
 \def\Pgy{\ifmmode\mathrm{\psi(2S)}
          \else$\mathrm{\psi(2S)}$\fi}
 \def\Pgya{\ifmmode\mathrm{\psi(3770)}
           \else$\mathrm{\psi(3770)}$\fi}
 \def\Pgyb{\ifmmode\mathrm{\psi(4040)}
           \else$\mathrm{\psi(4040)}$\fi}
 \def\Pgyc{\ifmmode\mathrm{\psi(4160)}
           \else$\mathrm{\psi(4160)}$\fi}
 \def\Pgyd{\ifmmode\mathrm{\psi(4415)}
           \else$\mathrm{\psi(4415)}$\fi}
 \def\PgU{\ifmmode\mathrm{\Upsilon(1S)}
          \else$\mathrm{\Upsilon(1S)}$\fi}
 \def\Pbgcz{\ifmmode\mathrm{{\chi}_{b0}(1P)}
            \else$\mathrm{{\chi}_{b0}(1P)}$\fi}
 \def\Pbgci{\ifmmode\mathrm{{\chi}_{b1}(1P)}
            \else$\mathrm{{\chi}_{b1}(1P)}$\fi}
 \def\Pbgcii{\ifmmode\mathrm{{\chi}_{b2}(1P)}
             \else$\mathrm{{\chi}_{b2}(1P)}$\fi}
 \def\PgUa{\ifmmode\mathrm{\Upsilon(2S)}
           \else$\mathrm{\Upsilon(2S)}$\fi}
 \def\Pbgcza{\ifmmode\mathrm{{\chi}_{b0}(2P)}
             \else$\mathrm{{\chi}_{b0}(2P)}$\fi}
 \def\Pbgcia{\ifmmode\mathrm{{\chi}_{b1}(2P)}
             \else$\mathrm{{\chi}_{b1}(2P)}$\fi}
 \def\Pbgciia{\ifmmode\mathrm{{\chi}_{b2}(2P)}
              \else$\mathrm{{\chi}_{b2}(2P)}$\fi}
 \def\PgUb{\ifmmode\mathrm{\Upsilon(3S)}
           \else$\mathrm{\Upsilon(3S)}$\fi}
 \def\PgUc{\ifmmode\mathrm{\Upsilon(4S)}
           \else$\mathrm{\Upsilon(4S)}$\fi}
 \def\PgUd{\ifmmode\mathrm{\Upsilon(10860)}
           \else$\mathrm{\Upsilon(10860)}$\fi}
 \def\PgUe{\ifmmode\mathrm{\Upsilon(11020)}
           \else$\mathrm{\Upsilon(11020)}$\fi}
 \def\Pp{\ifmmode\mathrm{p}
         \else$\mathrm{p}$\fi}
 \def\Pap{\ifmmode\mathrm{\overline{p}}
         \else$\mathrm{\overline{p}}$\fi}
 \def\Pn{\ifmmode\mathrm{n}
         \else$\mathrm{n}$\fi}
 \def\PNa{\ifmmode\mathrm{N(1440)P_{11}}
          \else$\mathrm{N(1440)P_{11}}$\fi}
 \def\PNb{\ifmmode\mathrm{N(1520)D_{13}}
          \else$\mathrm{N(1520)D_{13}}$\fi}
 \def\PNc{\ifmmode\mathrm{N(1535)S_{11}}
          \else$\mathrm{N(1535)S_{11}}$\fi}
 \def\PNd{\ifmmode\mathrm{N(1650)S_{11}}
          \else$\mathrm{N(1650)S_{11}}$\fi}
 \def\PNe{\ifmmode\mathrm{N(1675)D_{15}}
          \else$\mathrm{N(1675)D_{15}}$\fi}
 \def\PNf{\ifmmode\mathrm{N(1680)F_{15}}
          \else$\mathrm{N(1680)F_{15}}$\fi}
 \def\PNg{\ifmmode\mathrm{N(1700)D_{13}}
          \else$\mathrm{N(1700)D_{13}}$\fi}
 \def\PNh{\ifmmode\mathrm{N(1710)P_{11}}
          \else$\mathrm{N(1710)P_{11}}$\fi}
 \def\PNi{\ifmmode\mathrm{N(1720)P_{13}}
          \else$\mathrm{N(1720)P_{13}}$\fi}
 \def\PNj{\ifmmode\mathrm{N(2190)G_{17}}
          \else$\mathrm{N(2190)G_{17}}$\fi}
 \def\PNk{\ifmmode\mathrm{N(2220)H_{19}}
          \else$\mathrm{N(2220)H_{19}}$\fi}
 \def\PNl{\ifmmode\mathrm{N(2250)G_{19}}
          \else$\mathrm{N(2250)G_{19}}$\fi}
 \def\PNm{\ifmmode\mathrm{N(2600)I_{1,11}}
          \else$\mathrm{N(2600)I_{1,11}}$\fi}
 \def\PgDa{\ifmmode\mathrm{\Delta(1232)P_{33}}
           \else$\mathrm{\Delta(1232)P_{33}}$\fi}
 \def\PgDb{\ifmmode\mathrm{\Delta(1620)S_{31}}
           \else$\mathrm{\Delta(1620)S_{31}}$\fi}
 \def\PgDc{\ifmmode\mathrm{\Delta(1700)D_{33}}
           \else$\mathrm{\Delta(1700)D_{33}}$\fi}
 \def\PgDd{\ifmmode\mathrm{\Delta(1900)S_{31}}
           \else$\mathrm{\Delta(1900)S_{31}}$\fi}
 \def\PgDe{\ifmmode\mathrm{\Delta(1905)F_{35}}
           \else$\mathrm{\Delta(1905)F_{35}}$\fi}
 \def\PgDf{\ifmmode\mathrm{\Delta(1910)P_{31}}
           \else$\mathrm{\Delta(1910)P_{31}}$\fi}
 \def\PgDh{\ifmmode\mathrm{\Delta(1920)P_{33}}
           \else$\mathrm{\Delta(1920)P_{33}}$\fi}
 \def\PgDi{\ifmmode\mathrm{\Delta(1930)D_{35}}
           \else$\mathrm{\Delta(1930)D_{35}}$\fi}
 \def\PgDj{\ifmmode\mathrm{\Delta(1950)F_{37}}
           \else$\mathrm{\Delta(1950)F_{37}}$\fi}
 \def\PgDk{\ifmmode\mathrm{\Delta(2420)H_{3,11}}
           \else$\mathrm{\Delta(2420)H_{3,11}}$\fi}
 \def\PgDpp{\ifmmode\mathrm{\Delta^{++}}
           \else$\mathrm{\Delta^{++}}$\fi}
 \def\PgL{\ifmmode\mathrm{\Lambda}
          \else$\mathrm{\Lambda}$\fi}
 \def\PagL{\ifmmode\mathrm{\overline{\Lambda}}
            \else$\mathrm{\overline{\Lambda}}$\fi}
 \def\PgLa{\ifmmode\mathrm{\Lambda(1405) S_{01}}
           \else$\mathrm{\Lambda(1405) S_{01}}$\fi}
 \def\PgLb{\ifmmode\mathrm{\Lambda(1520) D_{03}}
           \else$\mathrm{\Lambda(1520) D_{03}}$\fi}
 \def\PgLc{\ifmmode\mathrm{\Lambda(1600) P_{01}}
           \else$\mathrm{\Lambda(1600) P_{01}}$\fi}
 \def\PgLd{\ifmmode\mathrm{\Lambda(1670) S_{01}}
           \else$\mathrm{\Lambda(1670) S_{01}}$\fi}
 \def\PgLe{\ifmmode\mathrm{\Lambda(1690) D_{03}}
           \else$\mathrm{\Lambda(1690) D_{03}}$\fi}
 \def\PgLf{\ifmmode\mathrm{\Lambda(1800) S_{01}}
           \else$\mathrm{\Lambda(1800) S_{01}}$\fi}
 \def\PgLg{\ifmmode\mathrm{\Lambda(1810) P_{01}}
           \else$\mathrm{\Lambda(1810) P_{01}}$\fi}
 \def\PgLh{\ifmmode\mathrm{\Lambda(1820) F_{05}}
           \else$\mathrm{\Lambda(1820) F_{05}}$\fi}
 \def\PgLi{\ifmmode\mathrm{\Lambda(1830) D_{05}}
           \else$\mathrm{\Lambda(1830) D_{05}}$\fi}
 \def\PgLj{\ifmmode\mathrm{\Lambda(1890) P_{03}}
           \else$\mathrm{\Lambda(1890) P_{03}}$\fi}
 \def\PgLk{\ifmmode\mathrm{\Lambda(2100) G_{07}}
           \else$\mathrm{\Lambda(2100) G_{07}}$\fi}
 \def\PgLl{\ifmmode\mathrm{\Lambda(2110) F_{05}}
           \else$\mathrm{\Lambda(2110) F_{05}}$\fi}
 \def\PgLm{\ifmmode\mathrm{\Lambda(2350) H_{09}}
           \else$\mathrm{\Lambda(2350) H_{09}}$\fi}
 \def\PgS{\ifmmode{\rm \Sigma}
           \else${\rm \Sigma}$\fi}
 \def\PgSp{\ifmmode\mathrm{\Sigma^+}
           \else$\mathrm{\Sigma^+}$\fi}
 \def\PgSz{\ifmmode\mathrm{\Sigma^0}
           \else$\mathrm{\Sigma^0}$\fi}
 \def\PgSm{\ifmmode\mathrm{\Sigma^-}
           \else$\mathrm{\Sigma^-}$\fi}
 \def\PgSpm{\ifmmode\mathrm{\Sigma^{\pm}}
           \else$\mathrm{\Sigma^{\pm}}$\fi}
 \def\PgSa{\ifmmode\mathrm{\Sigma(1385) P_{13}}
           \else$\mathrm{\Sigma(1385) P_{13}}$\fi}
 \def\PgSb{\ifmmode\mathrm{\Sigma(1660) P_{11}}
           \else$\mathrm{\Sigma(1660) P_{11}}$\fi}
 \def\PgSc{\ifmmode\mathrm{\Sigma(1670) D_{13}}
           \else$\mathrm{\Sigma(1670) D_{13}}$\fi}
 \def\PgSd{\ifmmode\mathrm{\Sigma(1750) S_{11}}
           \else$\mathrm{\Sigma(1750) S_{11}}$\fi}
 \def\PgSe{\ifmmode\mathrm{\Sigma(1775) D_{15}}
           \else$\mathrm{\Sigma(1775) D_{15}}$\fi}
 \def\PgSf{\ifmmode\mathrm{\Sigma(1915) F_{15}}
           \else$\mathrm{\Sigma(1915) F_{15}}$\fi}
 \def\PgSg{\ifmmode\mathrm{\Sigma(1940) D_{13}}
           \else$\mathrm{\Sigma(1940) D_{13}}$\fi}
 \def\PgSh{\ifmmode\mathrm{\Sigma(2030) F_{17}}
           \else$\mathrm{\Sigma(2030) F_{17}}$\fi}
 \def\PgSi{\ifmmode\mathrm{\Sigma(2050)}
           \else$\mathrm{\Sigma(2050)}$\fi}
 \def\PgXz{\ifmmode\mathrm{\Xi^0}
           \else$\mathrm{\Xi^0}$\fi}
 \def\PgXm{\ifmmode\mathrm{\Xi^-}
           \else$\mathrm{\Xi^-}$\fi}
 \def\PgXa{\ifmmode\mathrm{\Xi(1530)}
           \else$\mathrm{\Xi(1530)}$\fi}
 \def\PgXas{\ifmmode\mathrm{\Xi(1530)P_{13}}
           \else$\mathrm{\Xi(1530)P_{13}}$\fi}
 \def\PgXb{\ifmmode\mathrm{\Xi(1690)}
           \else$\mathrm{\Xi(1690)}$\fi}
 \def\PgXbb{\ifmmode\mathrm{\Xi(1620)}
           \else$\mathrm{\Xi(1620)}$\fi}
 \def\PgXc{\ifmmode\mathrm{\Xi(1820)D_{13}}
           \else$\mathrm{\Xi(1820)D_{13}}$\fi}
 \def\PgXcs{\ifmmode\mathrm{\Xi(1820)}
           \else$\mathrm{\Xi(1820)}$\fi}
 \def\PgXd{\ifmmode\mathrm{\Xi(1950)}
           \else$\mathrm{\Xi(1950)}$\fi}
 \def\PgXe{\ifmmode\mathrm{\Xi(2030)}
           \else$\mathrm{\Xi(2030)}$\fi}
 \def\PgOm{\ifmmode\mathrm{\Omega^-}
           \else$\mathrm{\Omega^-}$\fi}
 \def\PgO{\ifmmode\mathrm{\Omega}
           \else$\mathrm{\Omega}$\fi}
 \def\PgOma{\ifmmode\mathrm{\Omega(2250)^-}
            \else$\mathrm{\Omega(2250)^-}$\fi}
 \def\PcgL{\ifmmode\mathrm{\Lambda_c}
            \else$\mathrm{\Lambda_c}$\fi}
 \def\PacgL{\ifmmode\mathrm{\overline{\Lambda}_c}
            \else$\mathrm{\overline{\Lambda}_c}$\fi}
 \def\PcgLp{\ifmmode\mathrm{\Lambda_c^+}
            \else$\mathrm{\Lambda_c^+}$\fi}
 \def\PcgLm{\ifmmode{\rm \Lambda_c^-}
            \else${\rm \Lambda_c^-}$\fi}
 \def\PcgX{\ifmmode\mathrm{\Xi_c}
            \else$\mathrm{\Xi_c}$\fi}
 \def\PcgXz{\ifmmode\mathrm{\Xi_c^0}
            \else$\mathrm{\Xi_c^0}$\fi}
 \def\PcgXp{\ifmmode\mathrm{\Xi_c^+}
            \else$\mathrm{\Xi_c^+}$\fi}
 \def\PcgS{\ifmmode\mathrm{\Sigma_c}
           \else$\mathrm{\Sigma_c}$\fi}
 \def\PcgSz{\ifmmode\mathrm{\Sigma_c^0}
           \else$\mathrm{\Sigma_c^0}$\fi}
 \def\PcgSp{\ifmmode\mathrm{\Sigma_c^+}
           \else$\mathrm{\Sigma_c^+}$\fi}
 \def\PcgSpp{\ifmmode\mathrm{\Sigma_c^{++}}
           \else$\mathrm{\Sigma_c^{++}}$\fi}
 \def\PcgO{\ifmmode{\mathrm \Omega_c}
           \else${\mathrm \Omega_c}$\fi}
 \def\PcgOz{\ifmmode{\mathrm \Omega_c^{0}}
           \else${\mathrm \Omega_c^{0}}$\fi}
 \def\PSgg{\ifmmode\mathrm{\tilde{\gamma}}
           \else$\mathrm{\tilde{\gamma}}$\fi}
 \def\PSgxz{\ifmmode\mathrm{\tilde{\chi}^0_i}
            \else$\mathrm{\tilde{\chi}^0_i}$\fi}
 \def\PSZz{\ifmmode\mathrm{\tilde{Z}^0}
           \else$\mathrm{\tilde{Z}^0}$\fi}
 \def\PSHz{\ifmmode\mathrm{\tilde{H}^0_j}
           \else$\mathrm{\tilde{H}^0_j}$\fi}
 \def\PSgxpm{\ifmmode\mathrm{\tilde{\chi}^{\pm_i}}
             \else$\mathrm{\tilde{\chi}^{\pm_i}}$\fi}
 \def\PSWpm{\ifmmode\mathrm{\tilde{W}^{\pm}}
            \else$\mathrm{\tilde{W}^{\pm}}$\fi}
 \def\PSHpm{\ifmmode\mathrm{\tilde{H}^{\pm_j}}
            \else$\mathrm{\tilde{H}^{\pm_j}}$\fi}
 \def\PSgn{\ifmmode\mathrm{\tilde{\nu}}
           \else$\mathrm{\tilde{\nu}}$\fi}
 \def\PSe{\ifmmode\mathrm{\tilde{e}}
          \else$\mathrm{\tilde{e}}$\fi}
 \def\PSgm{\ifmmode\mathrm{\tilde{\mu}}
           \else$\mathrm{\tilde{\mu}}$\fi}
 \def\PSgt{\ifmmode\mathrm{\tilde{\tau}}
           \else$\mathrm{\tilde{\tau}}$\fi}
 \def\PSq{\ifmmode\mathrm{\tilde{q}}
          \else$\mathrm{\tilde{q}}$\fi}
 \def\PSg{\ifmmode\mathrm{\tilde{g}}
          \else$\mathrm{\tilde{g}}$\fi}
\begin{document}
\title{PENTAQUARKS - FACTS AND MYSTERIES\\
or SISYPHUS AT WORK}

\author{
Josef Pochodzalla\\
{\em Institut f\"ur Kernphysik, Universit\"at Mainz} \\
}
\maketitle
\baselineskip=11.6pt
\begin{abstract}
Recent evidence for pentaquark baryons is critically reviewed in
the light of new high statistics data. The search of the WA89
experiment for the $\Xi^{--}(1860)$ is presented in detail and
consequences of its non-observations are discussed.
\end{abstract}
\baselineskip=14pt
\section{The Myth of Sisyphus}
Giving these days a talk on pentaquarks or - even worse - writing
afterwards a report for the proceedings reminds very much on
Sisyphus, a man eternally condemned to roll a rock to the top of a
mountain, whence the stone would fall back to its own weight.
Having just finished the transparencies for the talk, the next
paper with a new -- positive or negative -- result  appears. In
that sense, the present manuscript written during june 2004
represents an updated version of the talk given at the PANDA
workshop in march 2004.

But may be there is even a deeper link between the pentaquark
search and the destiny of Sisyphus. Since its advent in 1964 the
quark model\cite{QUARK} is very much appreciated for describing
the vast amount of strongly interacting particles, the so called
hadron-zoo. Experimentally there is no doubt of the existence of
baryons, made up of three quarks, and mesons, consisting of a
quark anti-quark pair. A priori the quark model imposes no upper
limit on the number of quarks/anti-quarks a hadron can be built
of. However, it is widely agreed upon, that the colour quantum
numbers of the constituents should add up to the colour neutral
state. As a consequence physicist desperately seek for exotic
quark and gluon structures which differ from the well known meson
and baryon structure. Narrow resonances with exotic quark content
would be of course particularly welcome because the theoretical
interpretation would be very much simplified. In the past many new
particles have been spotted like the tetra quark
$U$(3100)\cite{WA62_U,BIS2_U}, the $f_J$(2230) seen first by the
MARK III collaboration\cite{fj2230_1} and the
$S$(1936)\cite{S1936}. Unfortunately none of these narrow
resonances survived detailed studies with high statistics. So here
we go again...

%--------------------------Fig 1 ------------------------------------------
\begin{figure}[t]
 \vspace{9.5cm}
\includegraphics{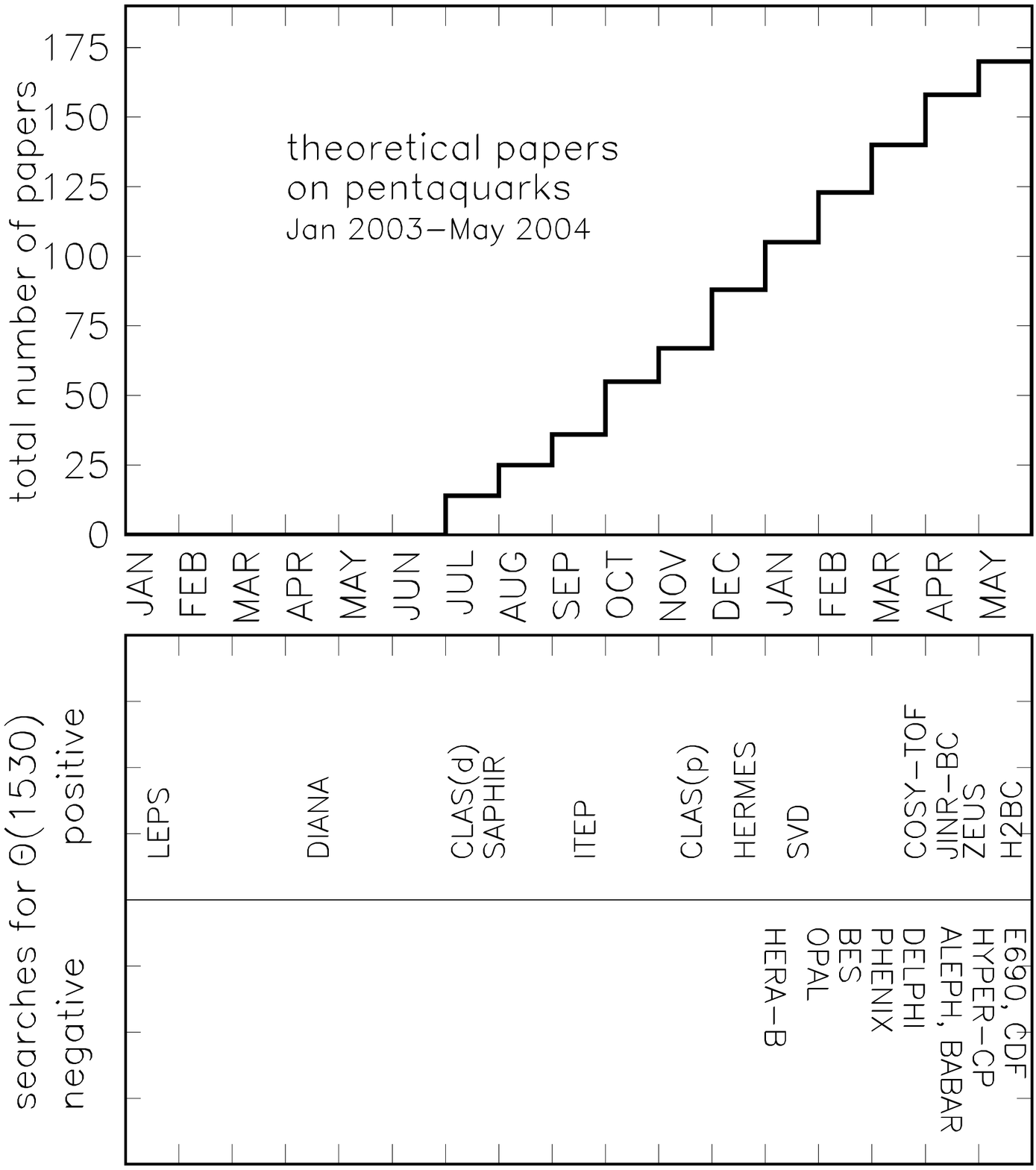}
 \caption{\it
      Evolution of the total number of manuscripts discussing
      pentaquarks during the last months (top) and experiments reporting on the
      observation or non-observation of the $\Theta^+(1530)$ signal. Please
      note that some of the experimental results are displaced in
      this plot despite the fact that they have been
      presented nearly simultaneously.
    \label{fig:WA8901} }
\end{figure}
%---------------------------Fig 1 ------------------------------------------

\section{The Experimental Situation of the $\Theta^+(1530)$}
At present twelve experimental groups have reported evidence for a
narrow baryonic resonance in the KN channel  at a mass of about
1530{\Mevc2} (see Refs. 6-17)
%\cite{Theta:LEPS,Theta:DIANA,Theta:CLASd,Theta:SAPHIR,Theta:CLASp,
%Theta:NEUTRINO,Theta:HERMES,Theta:SVD,Theta:COSY,Theta:bubble,Theta:ZEUS,Theta:JINRBC}
(for an updated list of references see\cite{Theta:Lit}). Based on
previous predictions\cite{Theta:Diakonov} (for some earlier
references see also\cite{Theta:Walliser}) this resonance was -
because of its exotic quark content - interpreted as a pentaquark
state. As a consequence already the first observations triggered a
flood of theoretical papers which is still increasing with an
increment of about one paper each second day (top part of Fig.
\ref{fig:WA8901}).
%---------------------------Fig 2 ---------------------------------------------
\begin{figure}[t]
 \vspace{7.5cm}
\includegraphics{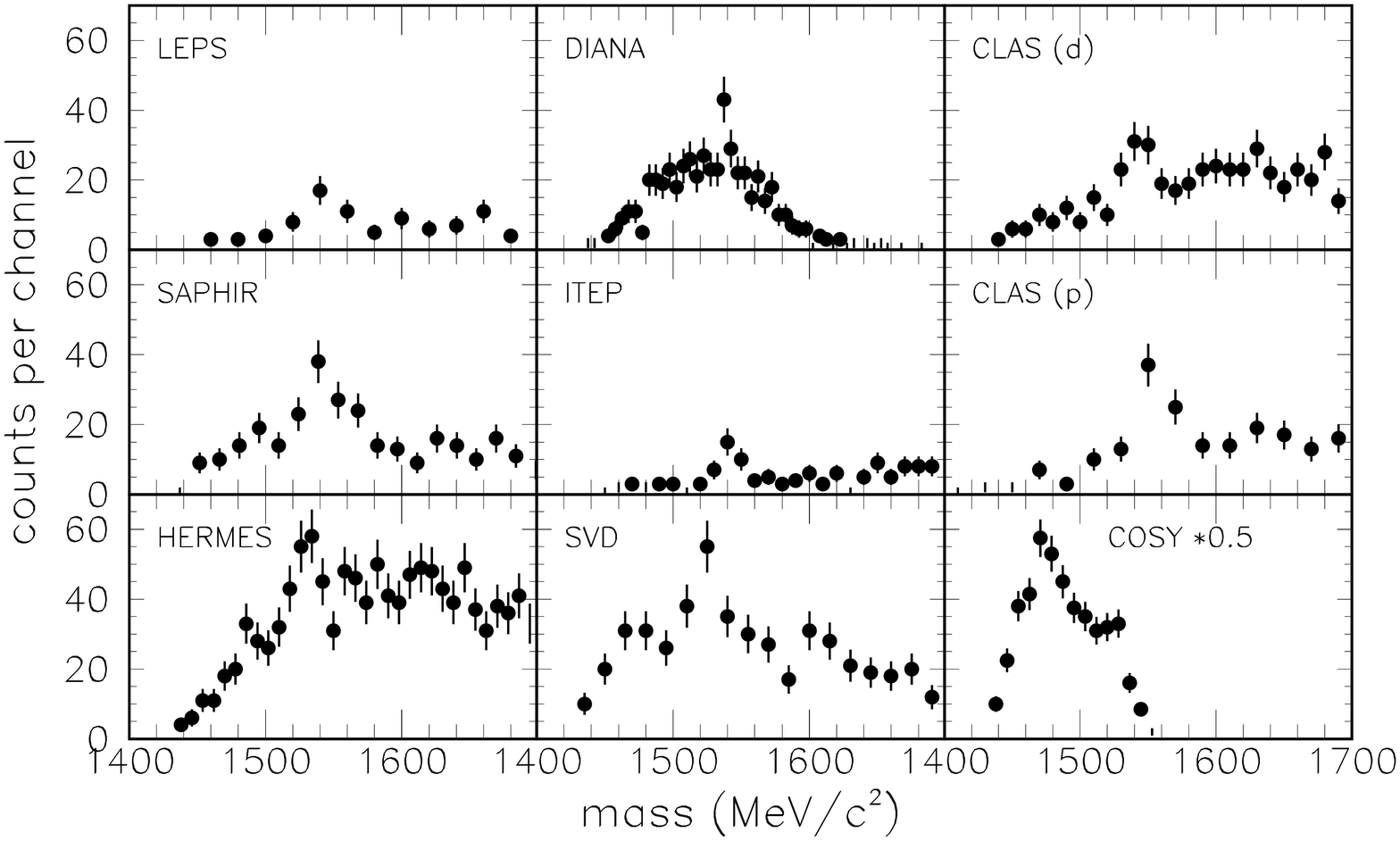}
 \caption{\it
      Summary of the first nine published observations of the $\Theta^{+}(1530)$
      resonance.
    \label{fig:WA8902} }
\end{figure}
%---------------------------Fig 2 ---------------------------------------------

Figure \ref{fig:WA8902} shows the first nine published results
which gave evidence for the existence of the so called
$\Theta^+(1530)$. Unlike in the original publications I prefer to
show here the data points including the statistical error bars.
Obviously a common drawback of the individual observations is the
limited statistics and hence limited confidence\cite{Bit00} of the
peaks. A little bit disturbing is also the fact that the magnitude
of the effect is nearly independent of the experimental situation.
Because of the low statistics it is important to note that any
cuts applied during the search process can modify the statistical
significance of an a priori unknown peak unless the cuts are
determined with an independent data sample or Monte Carlo data
(see e.g.\cite{Neeb}). The low statistics of the experiments shown
in Fig.~\ref{fig:WA8902} did usually not allow to separate the
data in two distinct data samples. It is furthermore interesting
that the position of the various peaks are not fully consistent.
Indeed already quite early doubts have been raised because of
possible experimental
artifacts\cite{Theta:Dzierba,Theta:Zavertyaev}. A recent analysis
of the HYPER-CP collaboration also underlines the necessity to
remove so called ghost tracks, i.e near-duplicate tracks, during
the analysis\cite{Theta:HYPERCP}. Using the positive track from a
$\Lambda$ decay twice as a $\pi^+$ and a proton produces a peak
near 1.54{\Gevc2} (cf. also the discussion on Fig.
\ref{fig:WA8907} below). Finally, even if the observed peaks were
real, more conventional processes can not be excluded completely
at the
moment\cite{Theta:Nussinov,Theta:Kahana,Theta:Kishimoto,Theta:Bicudo}
(see however Ref.\cite{Theta:Llanes}).

Since the beginning of this year also quite a number of negative
results became available (see lower part of
Fig.~\ref{fig:WA8901}). No signals of the $\Theta^+(1530)$ could
be found by BES\cite{Theta:BES}, HERA-B\cite{Xi:HERAB},
OPAL\cite{Theta:OPAL}, PHENIX\cite{Theta:PHENIX},
DELPHI\cite{Theta:DELPHI}, ALEPH\cite{Theta:ALEPH},
HYPER-CP\cite{Theta:HYPERCP}, E690\cite{Theta:E690},
CDF\cite{Theta:CDF} and BABAR\cite{Theta:BABAR}.  Although a
direct comparison of the positive and negative results is quite
difficult, the discovery potential of the various experiment can
be judged by the observed yield of known resonances. Whereas the
experiments with a positive result have -- if mentioned in the
publications at all -- typical $\Lambda(1520)$ yields of at most a
few hundred, the experiments with negative outcome report in
several cases a few thousand identified $\Lambda(1520)$ events. So
while counting naively just the number of reported results, the
situation is presently at near-balance (see
Fig.~\ref{fig:WA8901}), it seems that the critics have gained
already an advantage. It is therefore indisputable that further
high-statistics experiments are needed to establish the observed
resonance beyond any doubt. Once this has been achieved --
preliminary high statistics data of the LEPS collaboration seem to
confirm their first observation\cite{Theta:LEPSD} -- the
observation and non-observation of these resonance in different
reactions may help to shed some light on the production mechanism
and possibly also on the internal structure of these exotic
states.

% --------------------- FIGURE WA89 Fig 3 and 4 ------------------------------
\begin{figure}[t]
\begin{minipage}[t]{58mm}
\includegraphics[width=5.7cm]{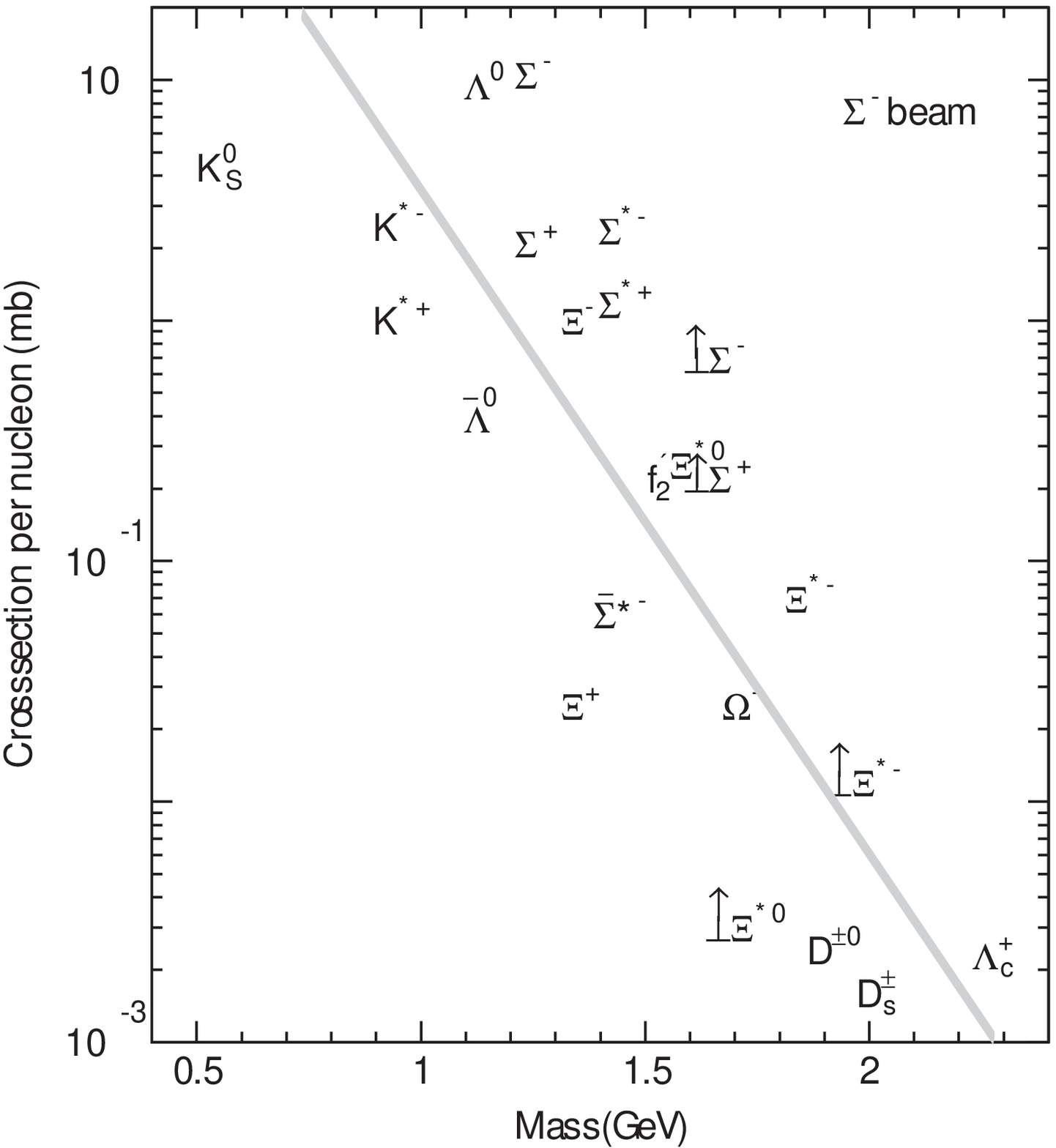}
\vspace{-0.0cm} \caption{\it Cross section per nucleon of various
strange and charmed hadrons observed by WA89 in $\Sigma^-$
reactions at 345 {\gevc1}.} \label{fig:WA8903}
\end{minipage}
\hspace{\fill}
\begin{minipage}[t]{58mm}
\includegraphics[width=5.2cm]{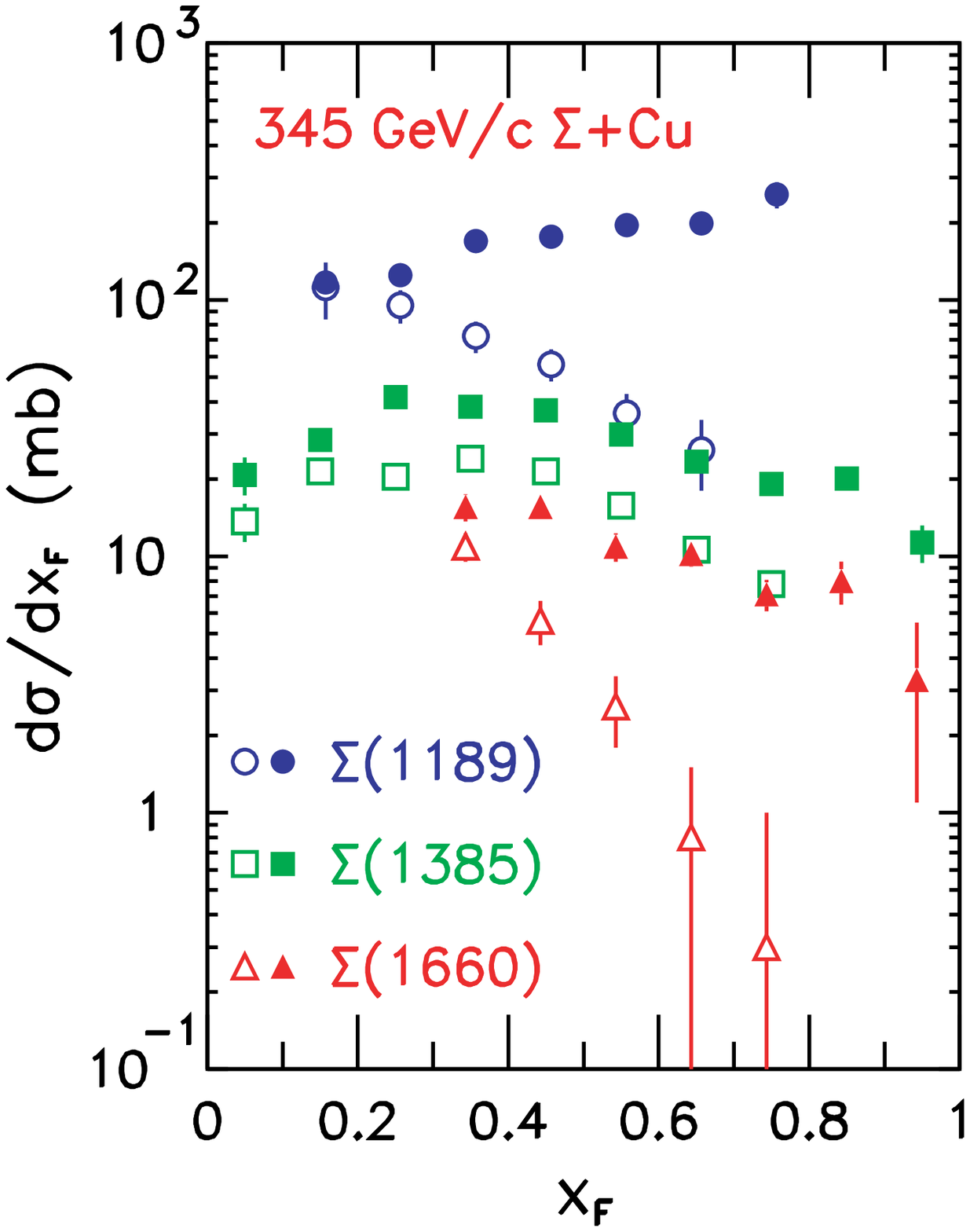}
\vspace{-0.0cm} \caption{\it $x_F$ distribution of positive (open
symbols) and negative (closed symbols) $\Sigma$ resonances studied
by WA89.}
\label{fig:WA8904}
\end{minipage}
\end{figure}

% --------------------- FIGURE WA89 fig 3 and 4 ------------------------------

\section{The $\Xi(1860)$ - Another Stone for Sisyphus?}
The interpretation of the observed peaks in terms of a five-quark
state was significantly strengthened by the subsequent observation
of another member of the anti\-cipated antidecuplet of
pentaquarks. Based on 1640 $\Xi^-$ candidates produced in p+p
interactions at 160{\gevc1} beam momentum, both in the
$\Xi^-\pi^+$ and the $\Xi^-\pi^-$ channels narrow peak structures
at an invariant mass of 1.860{\Gevc2} were observed by the NA49
collaboration\cite{Xi:NA49}. Possible signals of a $\Xi^*$
resonance at 1.860{\Gevc2} decaying into ${\Xi^-}\pi^+$ and
$Y\overline{K}$ were reported already 1977 for K$^-$p interactions
at 2.87{\gevc1}\cite{Bri77}. However, no corresponding signals
have been seen in other K$^-$ induced reactions (for a compilation
and a discussion of these data see Ref.\cite{Xi:Fischer}). A
preliminary analysis of proton-nucleus interactions at 920\gevc1
by the HERA-B collaboration using a total of 19000 reconstructed
$\Xi^-$ and $\overline{\Xi}^+$ events, shows no indication for the
$\Xi^{--}$ nor the $\Theta^+$ resonances\cite{Xi:HERAB}. Searches
for the $\Xi(1860)$ resonances are also being performed by the
ZEUS, CDF, ALEPH, E690 and the BABAR collaboration. The ZEUS data
comprise 1361 $\Xi^-$ and 1303 $\overline{\Xi}^+$ events, the CDF
sample contains 19150 $\Xi^-$ and 16736 $\overline{\Xi}^+$ and the
ALEPH collaboration collected about 1800 $\Xi^-$ . Negative --
though still preliminary -- results have been reported by all
three collaborations at the DIS04 conference\cite{Xi:CDFZEUS}. The
E690\cite{Theta:E690} and BABAR\cite{Xi:BABAR} experiments could
not find a significant signal despite a large data sample of
512000 and 258000 observed $\Xi^-$, respectively. First
preliminary results of the WA89 collaboration were presented at
the HYP03 conference already in october 2003\cite{HYP03}. The
final result presented in the following section are available in
Ref. \cite{Xi:WA89}

% --------------------- FIGURE WA89 Fig 5 and 6 ------------------------------
\begin{figure}[tb]
\begin{minipage}[t]{58mm}
\includegraphics[width=5.7cm]{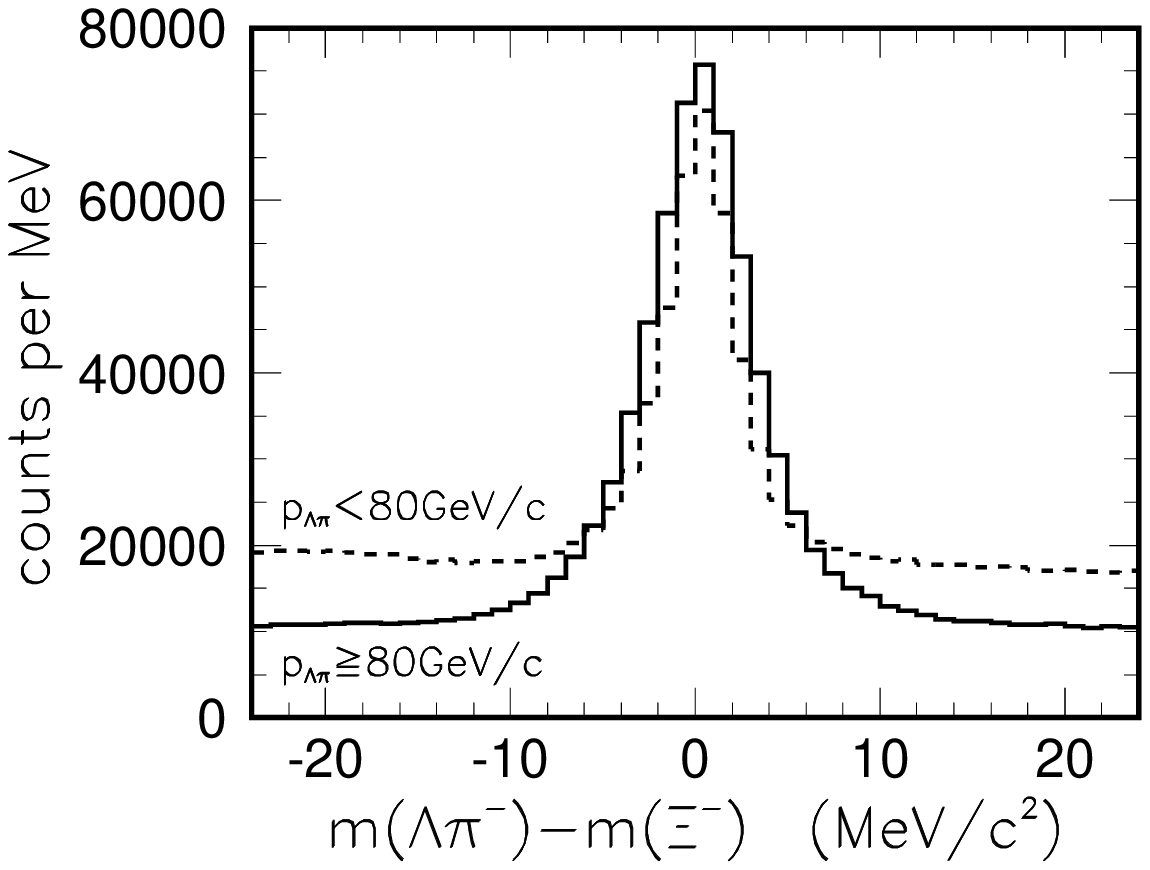}
\vspace{-0.0cm} \caption{\it Invariant mass distributions of
$\Lambda\pi^-$ pairs with p$_{\Lambda\pi} \geq$ 80{\gevc1} (solid
histogram) and $< $80{\gevc1} (dashed histogram) in 340{\gevc1}
$\Sigma^-$ induced interactions.}
\label{fig:WA8905}
\end{minipage}
\hspace{\fill}
\begin{minipage}[t]{58mm}
\includegraphics[width=5.7cm]{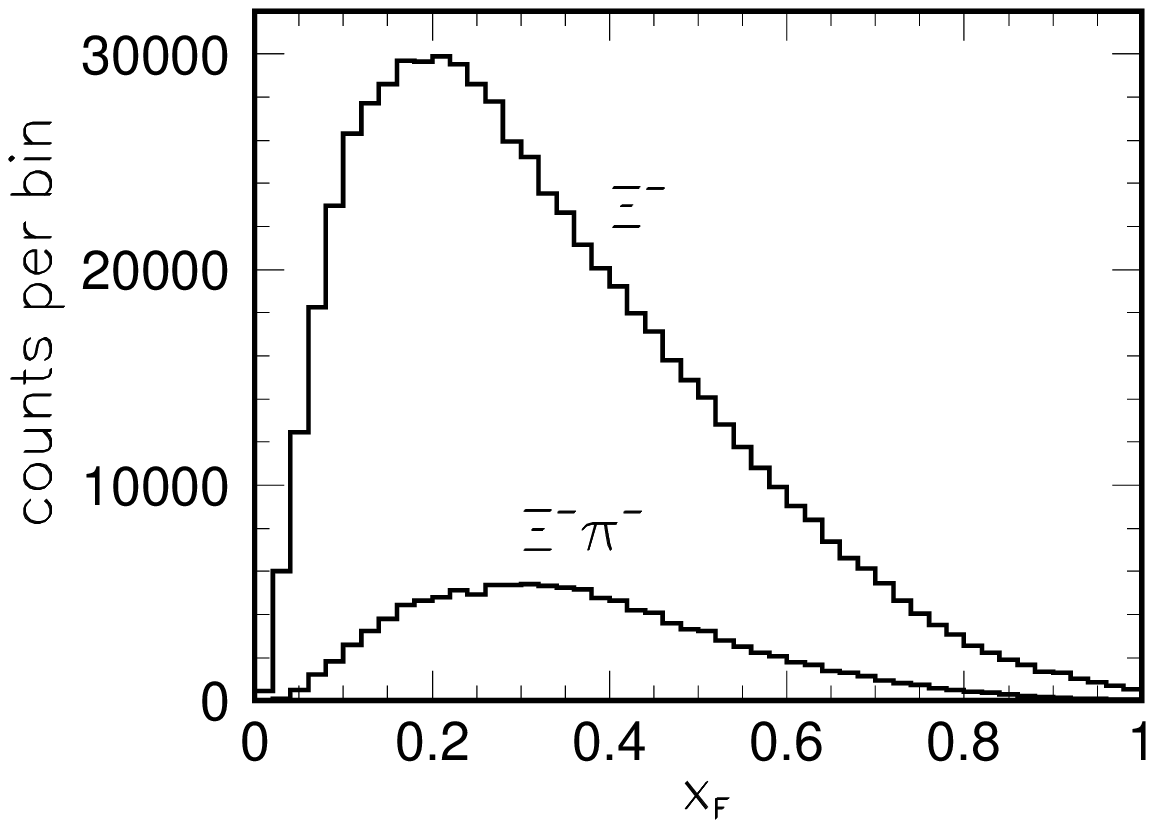}
\vspace{-0.0cm} \caption{\it Upper histogram: $x_F$ distribution
of the observed $\Xi^-$ events within a $\pm$2$\sigma$ mass
window. Lower histogram:  $x_F$ distribution of the observed
$\Xi^-\pi^-$ pairs within the mass range between 1.82 and 1.90
{\Gevc2}. In both cases the background has been subtracted by
means of sideband events.}
\label{fig:WA8906}
\end{minipage}
\end{figure}

% --------------------- FIGURE WA89 Fig 1+2 ------------------------------

\section{The Hyperon Beam Experiment WA89}

The hyperon beam experiment WA89 had the primary goal to study
charmed particles and their decays. At the same time it collected
a high statistics data sample of hyperons and hyperon
resonances\cite{{WA89:xi},{WA89:ximpip},{WA89:xistar},{WA89:sigma},{WA89:ll},{WA89:v0},{WA89:xipol}}.
The hyperon beamline\cite{WA89:beam} selected $\Sigma^-$ hyperons
with a mean momentum of 340~\gevc1 and a momentum spread of
$\sigma (p)/p=9\%$. In addition the beam contained small
admixtures of $K^-$ (2.1\%) and $\Xi^-$ (1.3\%)\cite{WA89:xi}. The
trajectories of incoming and outgoing particles were measured in
silicon microstrip detectors upstream and downstream of the
target. The experimental target itself consisted of one copper
slab with a thickness of 0.025 $\lambda_I$ in beam direction,
followed by three carbon (diamond powder) slabs of 0.008
$\lambda_I$ each, where $\lambda_I$ is the interaction length. The
momenta of the decay particles were measured in a magnetic
spectrometer equipped with MWPCs and drift chambers. In order to
allow hyperons and $K^0_S$ emerging from the target to decay in
front of the magnet the target was placed 13.6m upstream of the
center of the spectrometer magnet.

The symbols in Fig.~\ref{fig:WA8903} mark the cross sections per
nucleon for strange and charmed hadrons produced in $\Sigma^-$
induced reactions at 345{\gevc1}. In cases where the branching
ratio of the observed decay channel is not known only lower limits
are indicated by the vertical arrows. Typical for most hadronic
interactions in this energy regime the cross sections follow
roughly a mass dependence $\propto exp(-\Delta m/150MeV)$ as
indicated by the straight line.

The importance of the projectile for the hyperon production is
illustrated by the $x_F$ distributions of positive (open symbols)
and negative (closed symbols) $\Sigma$ resonances shown in
Fig.~\ref{fig:WA8904}. Whereas at large $x_F$ a significant
enhancement of negative hyperons of nearly a factor of 10 is
observed for the ground state, the decuplet resonance at 1385
\Mevc2 shows an enhancement of less than 3. Considering the fact
that for the $\Sigma^{\pm}_{1660}$ only values for $\sigma \cdot
BR$ are given, the large cross section for $\Sigma^-_{1660}$ seems
particularly striking (see closed triangles in
Fig.~\ref{fig:WA8904}). Furthermore, the $\Sigma^-_{1660}$ shows
again an enhancement over the $\Sigma^+_{1660}$ beyond a factor of
10. This is significantly larger than for $\Sigma_{1385}$ but
comparable to that of the ground state hyperons. Assuming that the
observed $\Sigma_{1660}$ is a $J^P=1/2^+$ octet state, the strong
leading effect for the $\Sigma_{1660}$ as compared to the rather
weak effect of the $\Sigma_{1385}$ decuplet may be related to the
$[ds]$ diquark structure. In the $J^P=3/2^+$ decuplet hyperon the
$[ds]$ diquarks have spin 1, while in the $\Sigma_{1660}$ the
$[ds]$ diquarks have predominantly spin 0.

\section{Search for the exotic $\Xi^{--}(1860)$ Resonance}

Since statistics is the key point when looking for new particles,
we also included interactions in the tracking detectors (silicon
detectors and plastic scintillator) located close to these targets
in our search for the S=-2 resonance in $\Sigma^-$ induced
reactions. $\Xi^-$ were reconstructed in the decay chain $\Xi^-
\rightarrow \Lambda\pi^- \rightarrow p\pi^-\pi^-$. The invariant
mass distributions of the $\Xi^-$ candidates are shown
Fig.~\ref{fig:WA8905} for two regions of the total momentum of the
$\Lambda\pi$ pair. The cut at 80{\gevc1} corresponds to an $x_F$
value of about 0.25 (see below). The WA89 analysis is based on a
total of 676k $\Xi^-$ candidates observed over a background of
170k $p\pi^-\pi^-$ combinations. Out of these candidates 240k,
281k and 155k can be attributed to the C, Cu and "Si+C+H" target,
respectively.

Because of the strangeness content of the $\Sigma^-$ beam also the
cross sections for $\Xi$ resonances are shifted towards large
$x_F$ with respect to the $\Sigma^-$-nucleon
cm-system\cite{WA89:xistar}. Since in the WA89 setup the
efficiency drops significantly at $x_F<$0.1 the yield of $\Xi^-$
peaks at $x_F \approx$ 0.2 (upper histogram in
Fig.~\ref{fig:WA8906}). $\Xi^-\pi^-$ pairs within the mass range
of 1.82 to 1.90 {\Gevc2} are shifted to even larger $x_F$ (lower
histogram in Fig.~\ref{fig:WA8906}). For comparison, the $\Xi^-$
events observed by NA49 are distributed over an $x_F$ range
between -0.25 and +0.25\cite{NA49:Barna}.

% --------------------- FIGURE WA89 Fig 7 + 8 ------------------------------

\begin{figure}[tb]
\begin{minipage}[b]{58mm}

\includegraphics[width=5.5cm]{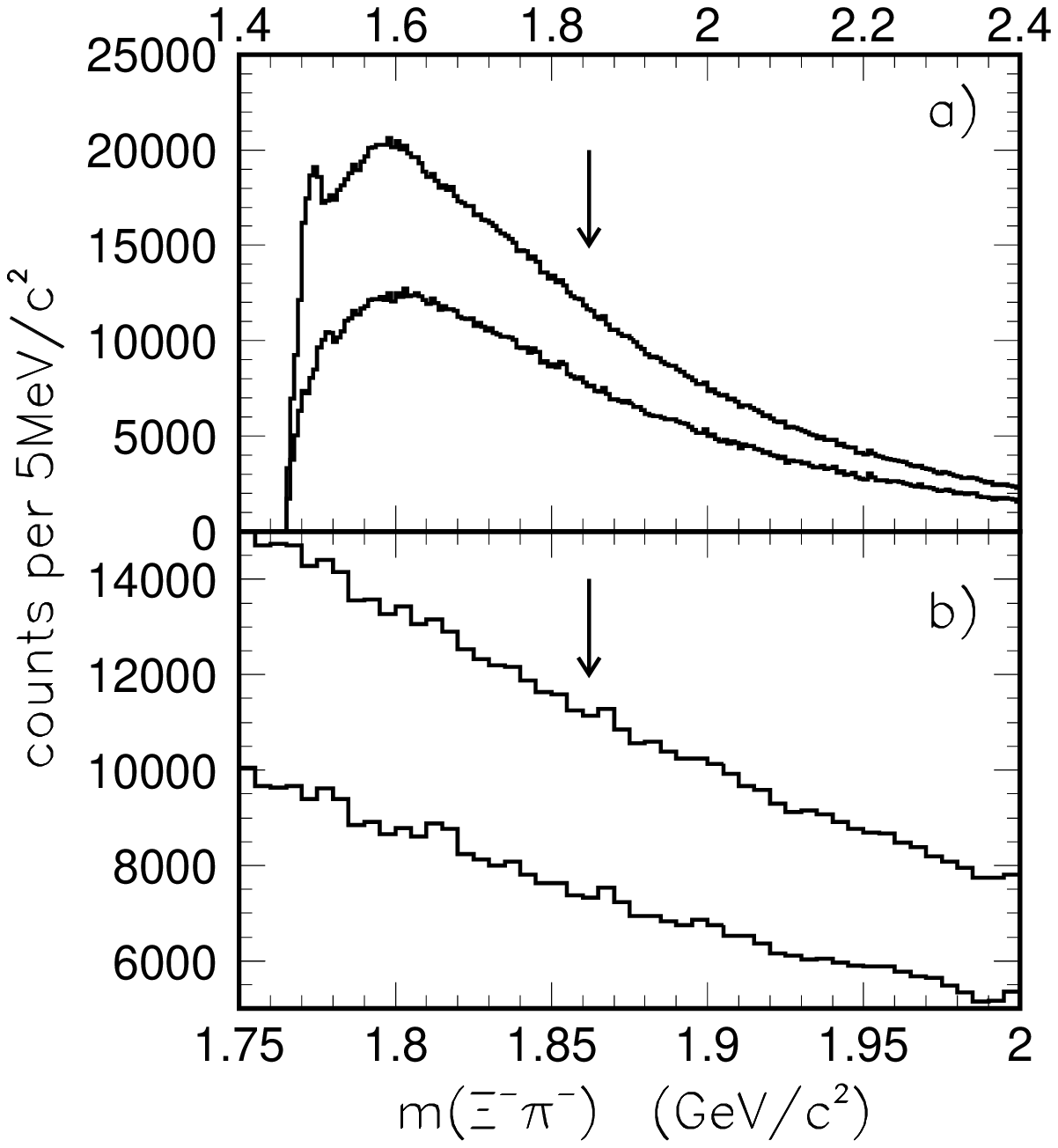}
\vspace{0.0cm} \caption{\it Effective mass distribution of
$\Xi^-\pi^-$ combinations of all reactions, including also
reactions in the tracking detectors (Si+C+H) close to the C and Cu
targets.  Part b) shows an extended view of the region around
1.862\Gevc2\ marked by the arrows. Note the offset of the y-axis
in this panel. In each panel the lower histogram shows the
distribution after background subtraction via sidebands.}
\label{fig:WA8907}
\end{minipage}
\hspace{\fill}
\begin{minipage}[b]{58mm}
\includegraphics[width=5.5cm]{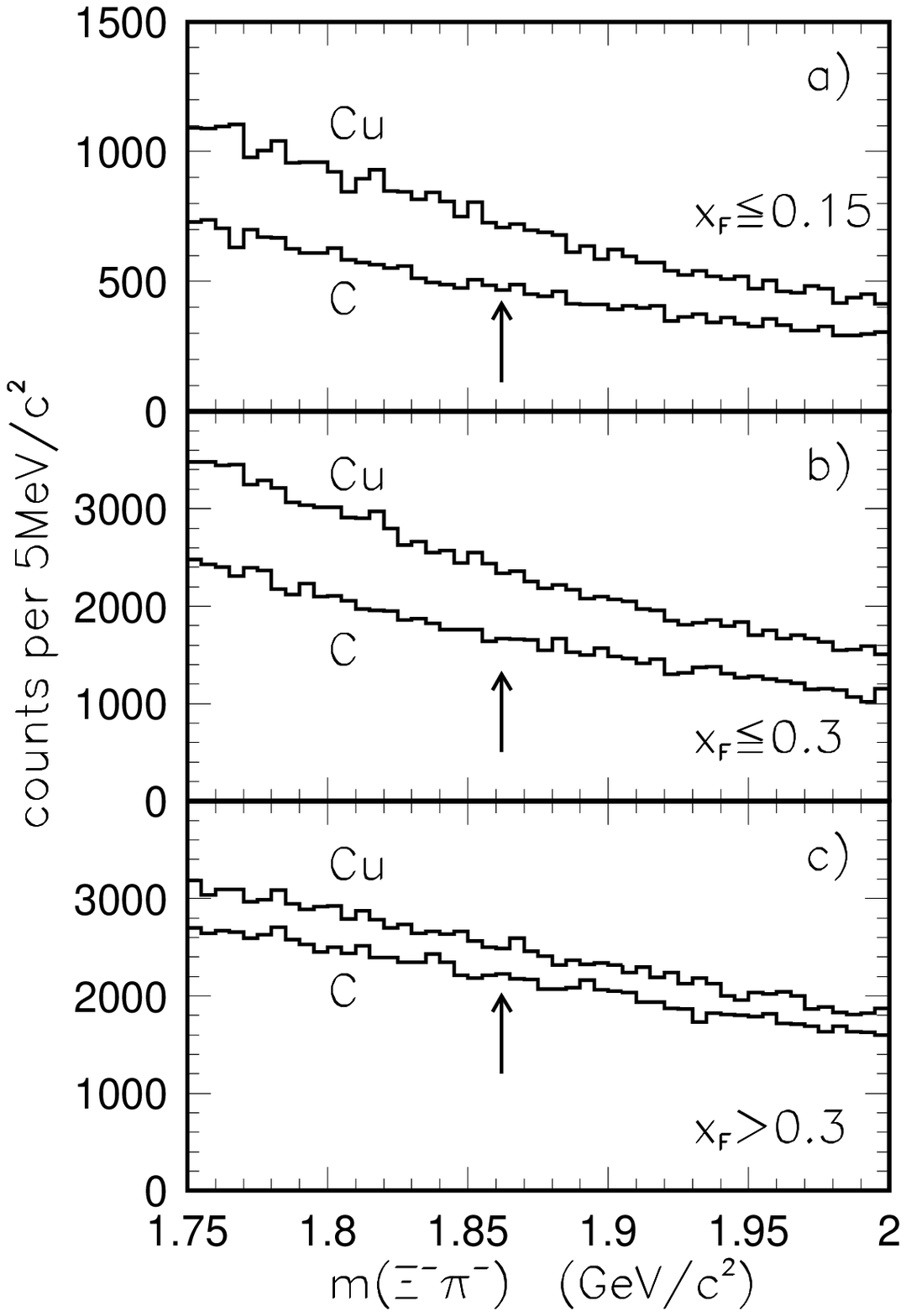}
\vspace{-0.0cm} \caption{\it Effective mass distribution of
$\Xi^-\pi^-$ combinations with $x_F(\Xi^-\pi^-)\leq$0.15 (part a),
$x_F(\Xi^-\pi^-)\leq$0.3 (part b) and $x_F(\Xi^-\pi^-) > $0.3
(part c). In each plot the lower and upper histogram correspond to
the carbon and copper target, respectively.} \label{fig:WA8908}
\end{minipage}
\end{figure}

% --------------------- FIGURE WA89 Fig 7 + 8 ------------------------------

Fig.~\ref{fig:WA8907} shows the invariant mass spectrum of all
observed $\Xi^-\pi^-$ pairs. Fig.~\ref{fig:WA8907}b shows an
extended view of the region around a mass of 1.862\Gevc2\ marked
by the arrows. All reactions, including also interactions in the
tracking detectors close to the C and Cu targets, contribute to
this figure. The structure observed at around 1.5\Gevc2 in the
upper histogram of Fig.~\ref{fig:WA8907}a is caused by events
where the negative pion from the decay of the $\Xi^-$ was wrongly
reconstructed as a double track. As can be seen from the lower
histogram in Fig.~\ref{fig:WA8907}a, these fake pairs are reduced
substantially by subtracting background from $\Xi^-$ sideband
events.

The NA49 collaboration has observed a ratio of $\Xi^{--}$ to
$\Xi^-$ candidates of about 1/40. If we assume the same {\it
relative} production cross sections over the full kinematic range
for the reaction in question and similar {\it relative} detection
efficiencies
$[{\varepsilon}(\Xi^{--})/{\varepsilon}(\Xi^-)]_{WA89}\approx
[{\varepsilon}(\Xi^{--})/{\varepsilon}(\Xi^-)]_{NA49}$ we would
expect of the order of 17000 $\Xi^{--} \rightarrow \Xi^-+\pi^-$
events in our full data sample. The FWHM of the peaks observed by
NA49 is 17\Mevc2 and is limited by the experimental resolution.
Since in our experiment the resolution is expected to be slightly
smaller $\approx$ 10{\Mevc2} (FWHM), this excess should be
concentrated in less than 6 channels in Fig.~\ref{fig:WA8907}b.
Obviously, no such enhancement can be seen in the spectra.

The $\Xi(1860)$ events observed by NA49 are concentrated at small
$x_F$. For a better comparison with the NA49 experiment we
therefore scanned our data for different ranges of $x_F$.
Fig.~\ref{fig:WA8908} shows the effective mass distributions of
$\Xi^-\pi^-$ combinations with $x_F(\Xi^-\pi^-)\leq$0.15,
$\leq$0.3 and $> $0.3 in the region around 1.862{\Gevc2}. In each
panel, the upper and lower histograms correspond to reactions with
the carbon and copper target, respectively. No background
subtraction was applied to these spectra.  Assuming again a
$\Xi^{--}$ to $\Xi^-$ ratio of 1/40 as observed by NA49 and
considering now only the $x_F$ range between 0 and 0.15, we
estimate that approximately 700 and 900 $\Xi^{--} \rightarrow
\Xi^-\pi^-$ events should be seen in Fig.~\ref{fig:WA8908}a for
the C and Cu target, respectively. None of these spectra shows
evidence for a statistically significant signal around
1.862{\Gevc2}, nor does such a signal appear in any other
sub-sample.

Upper limits on the production cross sections were estimated
separately for the copper and carbon targets, in five bins of
$x_F$ between $x_F=0.15$ and $x_F=0.9$. Assuming a dependence of
the cross section on the mass number as $\sigma_{nucl}\propto
\sigma_0\cdot A^{2/3}$, where $\sigma_0$ is the cross section {\em
per nucleon}, we obtained the limits on $BR\cdot d\sigma_0 /dx_F$.
Limits on the integrated production cross sections $\sigma$ were
then calculated by summing quadratically the contributions $
d\sigma /dx_F \cdot \Delta x_F$ in the five individual $x_F$ bins.
The results are $BR\cdot \sigma_{max}(0.15<x_F<0.9)$= 16 and 55
$\mu b$ per nucleus in case of the carbon and copper target,
respectively. An extrapolation to the cross sections per nucleon
yields the two values $BR\cdot \sigma_{0,max} = 3.1\, \mu b$ for
the carbon and $3.5\, \mu b$ for the copper target, in excellent
agreement with each other. As can be seen from
Fig.~\ref{fig:WA8903}, these limits do not exceed the production
cross sections of all other observed $\Xi^*$ resonances.

At large $x_F$ a significant fraction of the $\Xi^-$ are produced
by interactions induced by the $\Xi^-$ beam
contamination\cite{WA89:xi,WA89:xipol}. Even if we were to assume
that the $\Xi^{--}(1860)$ production can be attributed exclusively
to the 1.3\% $\Xi^-$ admixture in the beam, we obtain e.g. for the
carbon target and $x_F\geq$0.5 a limit for the $\Xi^{--}$
production by $\Xi^-$ of 740$\mu$b. For comparison, even this
large 3$\sigma$ limit corresponds to only 4\% of the $\Xi^-$
production cross section in $\Xi^-$+Be interactions at 116{\gevc1}
in the same kinematic range\cite{Bia87}.

% --------------------- FIGURE WA89 Fig 9 ------------------------------
\begin{figure}[t]
\vspace{5.0cm} \includegraphics{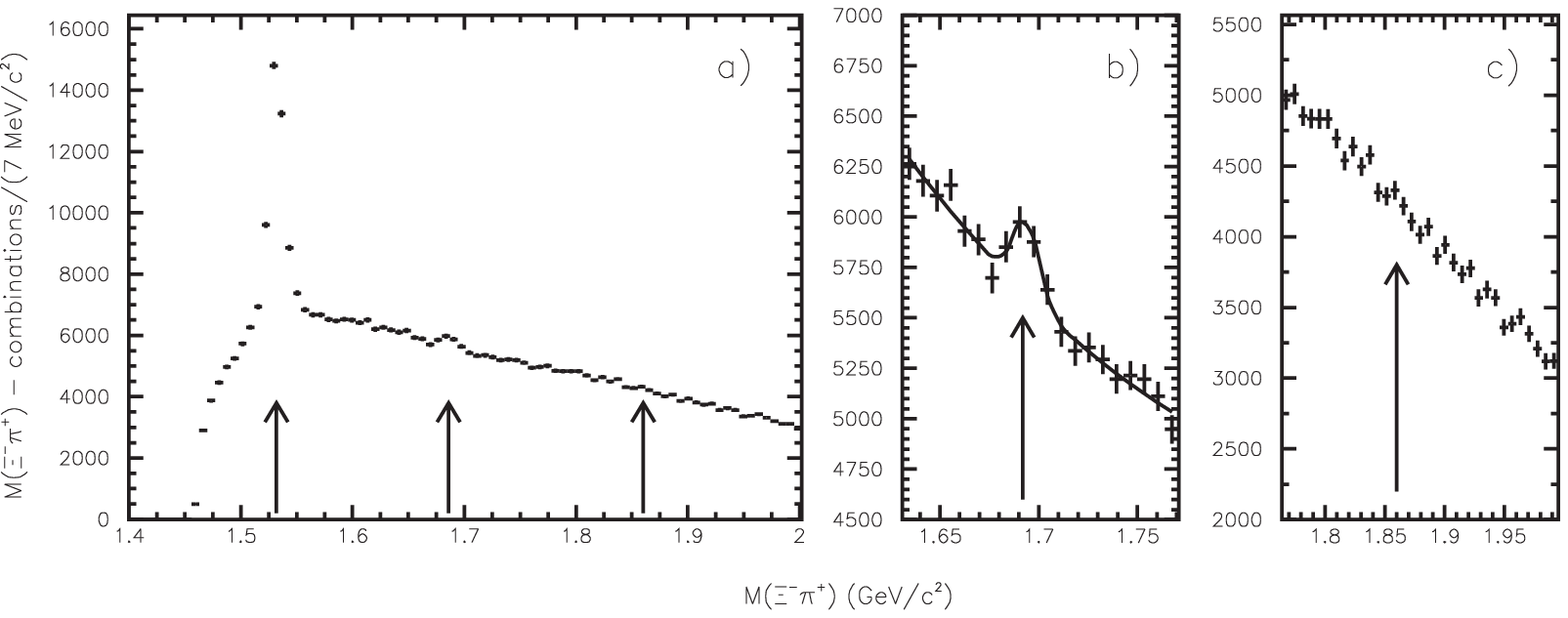}
\caption{\it Invariant mass distribution of the $\Xi^- \pi^+$
combinations published already six years ago\cite{WA89:ximpip}.
a)\ the $\Xi^{0}(1530)$ and $\Xi^{0}(1690)$ mass region; b) the
$\Xi^{0}(1690)$ mass region only; c) the mass region around
1860{\Mevc2}.
    \label{fig:WA8909} }
\end{figure}
%------------------------------------------------------------------------

Finally we note that the $\Xi^- \pi^+$ mass distribution observed
in the 1993 data set by WA89 has already been published some years
ago\cite{WA89:ximpip}. This combination is dominated by the peak
from $\Xi^0(1530)$ decays (see Fig.~\ref{fig:WA8909}). The
observed central mass was in good agreement with the known value
of M = 1531.8 $\pm$ 0.3{\Mevc2}\cite{pdg}. Unfolding the observed
width with the width of the $\Xi_{1530}$ of $\Gamma =
9.1${\Mevc2}\cite{pdg} gave an experimental resolution of $\sigma
_{\Xi^{0}(1530)}$ = 3.7{\Mevc2}. Furthermore, a weak resonance
signal with a width of $\Gamma = 10 \pm 6 \Mevc2\ $ is visible at
$M = 1686 \pm 4 \Mevc2\ $ above a large background. In the mass
region of the $\Xi^0(1860)$ no enhancement over the uncorrelated
background can be seen in the WA89 data.

\section{Quintessence}
After an euphoric stage with many favorable reports within a short
time we have now reached a phase which is much more unclear.
Counting just the number of reported results the situation of the
$\Theta^{+}(1530)$ is presently at most near-balance between
sightings and non-sightings. It seems, however, that -- because of
the higher statistics -- the non-sighting experiments gain the
preponderance.

Considering the seven non-observations of the $\Xi^{--}(1860)$
resonance compared to the single  claim in favor of it by the NA49
collaboration, this pentaquarks seems to stand of very shaky
ground at present. {\em If}, nonetheless, the $\Xi^{--}$ signal
observed by the NA49 collaboration is real, then the
non-observation in the WA89 experiment -- as well as the other
experiments -- is not easily understood. Generally particle ratios
do not vary significantly for the beam momentum range in question
(160\gevc1 vs. 340\gevc1) \cite{Let03,Liu04}. The fact that the
$\Theta^+(1530)$ has been seen in reactions on complex nuclei
\cite{Theta:NEUTRINO,Theta:SVD} makes also the different targets
(hydrogen vs. C, Si, Cu) an unlikely cause for the discrepancy.
The internal structure of the $\Sigma^-$ projectile or of the
$\Xi^{--}(1860)$ could be a more plausible reason for the rather
low limit of the $\Xi^{--}(1860)$/$\Xi^-$ ratio. It is well known,
that a transfer of a strange quark from the beam projectile to the
produced hadron enhances the production cross sections in
particular at large $x_F$ (see, for instance,
Fig.~\ref{fig:WA8904}). The different leading effects for octet
and decuplet  $\Sigma$ states\cite{WA89:sigma} even hint at an
$[sd]$ diquark transfer from the $\Sigma^-$
projectile\cite{WA89:poc01}. The production of a pentaquark
containing correlated quark-quark pairs (see e.g.
Ref.\cite{Jaf03}) would probably benefit from such a diquark
transfer. However, for example in case of an extended
$\overline{K}-N-\overline{K}$ molecular structure of the
$\Xi(1860)$\cite{Bic04} an $[sd]$ diquark transfer may not
necessarily enhance the $\Xi^{--}$ production leading also to a
narrower $x_F$ distribution. As a consequence the cross section in
$\Sigma^-$ induced reactions might not exceed the one for
production in pp interactions. The latter cross section is
predicted to be $\sim$ 4$\mu$b\cite{Liu04} which is then close to
our limit. Thus, if future high statistics experiments will
confirm the production of the $\Xi^{--}(1860)$ resonance in
proton-proton interaction, the non-observation with the $\Sigma^-$
beam would point to a very exceptional production mechanism
possibly related to an exotic structure of the $\Xi^{--}(1860)$.
However, the possible non-observation by the E690 collaboration
\cite{Theta:E690} in 800{\gevc1} p-p interactions may even ruin
this argument in favor of the $\Xi^{--}(1860)$ resonance.

Keeping in mind the past searches for exotic quark structures and
looking at the present contradictory data, we can therefore not
exclude that the stone of Sisyphus is just about to roll back
downhill and that the quest for exotic pentaquark states may end
where it began. May be QCD is indeed sticking to two and three
valence quarks only and may be we are just lacking the right
argument for this beautiful simplicity. In this situation it might
be helpful to recall what Albert Camus said about the poor
Sisyphus. Sisyphus {\it is} after all happy although he is fully
aware that he will not succeed:{\em The struggle itself toward the
heights is enough to fill a man's heart.}

%%%%%%%%%%%%%%%%%%%%%%%%%%%%%%%%%%%%%%%%%%%%%%%%%%%%%%%%%%%%%%%%%%%%%%
% The bibliography
%%%%%%%%%%%%%%%%%%%%%%%%%%%%%%%%%%%%%%%%%%%%%%%%%%%%%%%%%%%%%%%%%%%%%%

\end{document}